\documentclass[twocolumn,10pt,pre,groupedaddress,showpacs,preprintnumbers,amsmath,amssymb,twoside]{revtex4}

\usepackage{graphicx}
\usepackage{dcolumn}
\usepackage{bm}

\begin{document}

\title{Energy transfer in finite-size exciton-phonon systems : confinement-enhanced quantum decoherence} 
\author{Vincent Pouthier}
\email{vincent.pouthier@univ-fcomte.fr}
\affiliation{Institut UTINAM,  Universit\'{e} de Franche-Comt\'{e}, \\  CNRS UMR 6213, 25030 Besan\c {c}on Cedex, 
France}

\date{\today}

\begin{abstract}
Based on the operatorial formulation of the perturbation theory, the exciton-phonon problem is revisited for investigating 
exciton-mediated energy flow in a finite-size lattice. Within this method, the exciton-phonon entanglement is taken into 
account through a dual dressing mechanism so that exciton and phonons are treated on an equal footing. In a marked contrast with what happens in an infinite lattice, it is shown that the dynamics of the exciton density is governed by several time scales. The density evolves coherently in the short-time limit whereas a relaxation mechanism occurs over intermediated time scales. Consequently, in the long-time limit, the density converges toward a nearly uniform distributed equilibrium distribution. Such a behavior results from quantum decoherence that originates in the fact that the phonons evolve differently depending on the path followed by the exciton to tunnel along the lattice. Although the relaxation rate increases with the temperature and with the coupling, it decreases with the lattice size, suggesting that the decoherence is inherent to the confinement.
\end{abstract}


\maketitle

\section{Introduction}

In molecular lattices, understanding how excitons carry energy from one region to another is a key step for explaining many phenomena \cite{kn:may}. Example among many are amide-I excitons in $\alpha$-helices that favor the transduction of the chemical energy into mechanical work
\cite{kn:helix1,kn:helix2,kn:helix3,kn:edler,kn:helix4,kn:helix5,kn:helix6,kn:helix7,kn:helix8}, vibrons in adsorbed nanostructures that enhance surface reactions or promote quantum information transfer \cite{kn:surf1,kn:surf2,kn:surf3,kn:surf4,kn:QST}, and Frenkel excitons in light-harvesting complexes that convert solar energy into chemical energy \cite{kn:photo1,kn:photo2,kn:photo3,kn:photo4}. A molecular lattice exhibits regularly distributed atomic subunits along which the energy of an electronic transition, or of a high-frequency vibrational mode, delocalizes owing to dipole-dipole interaction. This gives rise to a narrow-band exciton whose eigenstates are superimpositions of local states. When the dynamics is governed by the exciton Hamiltonian, such superimpositions are coherent since a phase relation is kept between the local states. This favors a wavelike motion of the exciton that propagates coherently. 

Unfortunately, the exciton does not evolve freely but it interacts with the vibrations of the host medium usually responsible for dephasing. Therefore, the fundamental question arises whether the energy delocalizes coherently or incoherently \cite{kn:Scholes}. For instance, recent experiments revealed that the coherent nature of the exciton may explain the remarkable efficiency of light-harvesting complexes \cite{kn:Engel,kn:Calhoun}. Similarly, in a peptide helix, it has been shown that when a localized amide-I mode is selectively excited, the energy transport rate drastically increases \cite{kn:Backus} suggesting that a coherent energy flow takes place between amide-I modes \cite{kn:Kobus}. 

From a phenomenological point of view, dephasing limited coherent motion was described within stochastic models \cite{kn:stoch1,kn:stoch2,kn:stoch3,kn:stoch4,kn:stoch5,kn:stoch6}. The main idea is that the lattice vibrations induce random fluctuations of the site energies that behave as independent stochastic variables. Consequently, the phase of each local state randomly fluctuates so that the coherent nature of the exciton gradually disappears. A wavelike motion takes place in the short-time limit whereas a diffusion like motion occurs in the long-time limit. 
However, it turns out that the previous scenario is not so simple when a microscopic description is used to mimic the influence of the fluctuating surrounding. To proceed, most theories are based on either the Fr\"{o}hlich model \cite{kn:frohlich} or the Holstein model \cite{kn:holstein} that provide a general description of an exciton coupled with acoustic or optic phonons, respectively (see for instance Refs. \cite{kn:trans1,kn:trans2,kn:trans3,kn:trans4,kn:trans5,kn:trans6,kn:trans7,kn:trans8,kn:trans9,kn:pouthier8}). Depending on the model parameters, the exciton properties exhibit different facets ranging from quantum to classical, from weak coupling to strong coupling, from adiabatic to nonadiabatic and from large to small polarons \cite{kn:barisic}. 

In that context, a quite different result from that derived from stochastic approaches was obtained recently for a narrow-band exciton coupled with acoustic phonons in an infinite lattice. Indeed, within the nonadiabatic weak-coupling limit, that is, when the exciton moves slower than the phonons, it has been shown that the exciton propagates freely as if it was insensitive to the phonons \cite{kn:pouthier3}. This feature was established by using a time-convolutionless generalized master equation (TCL-GME) for describing the evolution of the exciton reduced density matrix (RDM) \cite{kn:book,kn:book1}.  
In an infinite lattice, the phonons behave as a reservoir insensitive to the exciton so that the Born approximation applies \cite{kn:barnett}. The influence of the phonons is encoded in a time-dependent relaxation operator. Up to second order, it involves exciton-phonon coupling correlation functions whose dynamics is governed by free phonons. Consequently, because acoustic phonons exhibit spatial correlations over an infinite length scale, the relaxation operator describes a fast dephasing-rephasing mechanism which prevents the exciton diffusivity. Note that similar results have been obtained for a donor-acceptor pair coupled with a solvent that exhibits spatial correlations \cite{kn:Nalbach}. 

In the present paper, the exciton-phonon problem is revisited for describing energy redistribution in a finite-size lattice, a question that is rarely addressed. However, characterizing size effects is of fundamental importance because in biology and in  nanoscience, relevant structures have reduced dimensionality. In a confined environment, we are faced with a major problem because the phonons no longer behave as a reservoir \cite{kn:esposito}. 
Indeed, as observed in the TCL-GME that governs the evolution of the excitonic coherences, that is, the off-diagonal RDM elements between the vacuum and the one-exciton states, the confinement favors quantum recurrences and strong memory effects in the exciton-phonon coupling correlation functions \cite{kn:pouthier4}. Therefore, the relaxation operator is an almost periodic function so that the TCL-GME reduces to a linear system of differential equations with almost periodic coefficients. In that case, parametric resonances give rise to an exponential growth of the RDM indicating that the TCL-GME method breaks down. The Born approximation fails to capture the dynamics because the correlations between the exciton and the phonons are no longer negligible during the propagation. 

To overcome these problems inherent in the non-Markovian dynamics, different strategies can be used, such as the correlated projector method \cite{kn:meth1}, the time-dependent projection-operator approach \cite{kn:meth2}, the effective-mode representation \cite{kn:meth3} and the quantum jumps approach \cite{kn:meth4}.
Here, we use the operatorial formulation of the perturbation theory (PT) \cite{kn:wagner} recently introduced for studying the dynamics of the excitonic coherences \cite{kn:pouthier5,kn:pouthier6,kn:pouthier7}. In the nonadiabatic weak-coupling limit, this method is a powerful tool for describing the spectral properties of the exciton-phonon system over a broad energy scale. It is particularly suitable for characterizing the coherence dynamics and is more accurate than TCL-GME. 
Within PT, the system dynamics is governed by an effective Hamiltonian that explicitly accounts for exciton-phonon entanglement. Exciton and phonons are thus treated on an equal footing to go beyond the Born approximation.

The paper is organized as follows. In Sec. II, the exciton-phonon Hamiltonian is described. Then, the results provided by PT are summarized to derive the effective exciton-phonon Hamiltonian. Finally, the exciton RDM is defined as the key observable required for characterizing the energy transfer. Its expression is established within PT. 
In Sec. III, the simulation of the RDM dynamics is carried out numerically to extract the time evolution of the exciton density. The results are finally discussed in Sec. IV.

\section{Theoretical background}

\subsection{Finite-size exciton-phonon system}

In a finite-size lattice containing $N$ sites $x=1,...,N$, the exciton-phonon Hamiltonian is defined as (see for instance Refs.\cite{kn:pouthier4,kn:pouthier5,kn:pouthier6,kn:pouthier7})
\begin{equation} 
H= H_A+H_B +\Delta H. 
\label{eq:H}
\end{equation}
In Eq.(\ref{eq:H}), the first term is the exciton Hamiltonian $H_A$ that acts in the one-exciton subspace $\mathcal{E}_A$. It refers to the dynamics of $N$ coupled two-level systems with Bohr frequency $\omega_0$. Note that a two-level system represents for instance the first two vibrational states of a localized high-frequency intramolecular vibration. The Hamiltonian $H_A$ is expressed in terms of the bare hopping constant $\Phi$ between nearest neighbor sites as (note that the convention $\hbar=1$ will be used throughout this paper) 
\begin{equation} 
H_A=\sum_{x=1}^{N} \omega_0 |x\rangle \langle x| + \sum_{x=1}^{N-1} \Phi(|x+1\rangle \langle x|+|x\rangle \langle x+1|), 
\label{eq:T}
\end{equation}
where $|x\rangle$ is the first excited state of the $x$th two-level system. Owing to the confinement, the exciton eigenstates are stationary waves with quantized wave vectors $K_{k}=k\pi/L$, with $k=1,..,N$ and $L=N+1$, as
\begin{equation} 
|k\rangle= \sum_{x=1}^{N}  \sqrt{\frac{2}{L}} \sin (K_kx) |x\rangle.
\label{eq:ket}
\end{equation}
The corresponding eigenenergies $\omega_{k}=\omega_0+2\Phi \cos(K_k)$ form a symmetric ladder of $N$ discrete energy levels that belong to a band centered on $\omega_0$ and whose width is approximately $4\Phi$. In the eigenbasis, the exciton Hamiltonian is thus defined as : $H_{A}= \sum_{k=1}^{N} \omega_k |k\rangle \langle k|$.

In Eq.(\ref{eq:H}), the second term is the phonon Hamiltonian $H_B=\sum_{p=1}^{N} \Omega_pa_p^{\dag}a_p$ that acts in the Hilbert space $\mathcal{E}_B$. It describes the external motions of the lattice sites that behave as point masses $M$ connected via force constants $W$. The phonons correspond to $N$ stationary normal modes with quantized wave vectors $q_{p}=p\pi/L$, with $p=1,..,N$. The corresponding frequencies are $\Omega_{p}=\Omega_{c} \sin(q_p/2)$ with $\Omega_{c}=\sqrt{4W/M}$. The dynamics is described using standard phonon operators $a_{p}^{\dag}$ and $a_{p}$ and the eigenstates are the well-known number states denoted $|n_p\rangle\equiv|n_1,...,n_N\rangle$. 

The last term in Eq.(\ref{eq:H}) stands for the exciton-phonon interaction $\Delta H=\sum_{p=1}^{N}M_p(a_p^{\dag}+a_p)$ that acts in $\mathcal{E}_A\otimes\mathcal{E}_B$. It is defined in terms of the coupling matrix $M_p$ that measures the strength of the interaction between the exciton and the $p$th phonon mode. The interaction yields a stochastic modulation of each two-level system Bohr frequency by the lattice vibrations. In the exciton eigenbasis, the matrix elements of $M_p$ are expressed as 
\begin{equation}
M_{pkk'} = \eta_p(\delta_{p,k-k'}+\delta_{p,k'-k}-\delta_{p,k+k'}-\delta_{p,2L-k-k'}),
\label{eq:Mk}
\end{equation}
where $\eta_p = [(E_B \Omega_p /L) ( 1- (\Omega_p/\Omega_c)^2 )]^{1/2}$ is expressed in terms of the small polaron binding energy $E_B=\chi^2/W$ that involves the coupling strength $\chi$. 
Equation (\ref{eq:Mk}) shows that $\Delta H$ favors exciton scattering from state $|k\rangle$ to state $|k'\rangle$ via the exchange of a phonon $p$. The allowed transitions are specified by the selection rules $M_{pkk'}\neq0$ that generalize the concept of momentum conservation in a finite-size lattice. 

The Hamiltonian $H$ will be used for characterizing exciton-mediated energy redistribution. To proceed, we shall focus our attention to the nonadiabatic ($4\Phi \ll \Omega_c$) weak-coupling  ($E_B\ll\Phi$) limit, a common situation for vibrational excitons in molecular lattices. In that case, $\Delta H$ being a small perturbation, the unperturbed states $|k, n_p \rangle$ refer to a free exciton accompanied by free phonons. Although $\Delta H$ favors transitions between unperturbed states, the key point is that there is no resonance between coupled unperturbed states so that PT can be used to solve $H$. 

\subsection{Perturbation theory}

The operatorial formulation of PT is based on the introduction of a unitary transformation $U$ that diagonalizes the Hamiltonian $\hat{H}=UHU^{\dag}$ in the unperturbed basis \cite{kn:pouthier5,kn:pouthier6}. It is written as $U=\exp(S)$, where $S$ is an anti-Hermitian generator that is non-diagonal in the unperturbed basis. Within PT, $S$ is expanded as a Taylor series with respect to $\Delta H$ so that the diagonalization is achieved at a given order (see appendix A). Up to second order,  the transformed Hamiltonian is written as 
\begin{equation}
\hat{H} = \sum_{k=1}^{N} \left[ (\omega_k+\delta \omega_k)|k\rangle \langle k|+ \hat{H}_B^{(k)} \otimes |k\rangle \langle k| \right],
\label{eq:Hhat}
\end{equation}
where $\hat{H}_B^{(k)}$ is the Hamiltonian that governs the phonon dynamics when the exciton lies in the state $|k\rangle$, as 
\begin{equation}
\hat{H}_B^{(k)}= \sum_{p=1}^{N} (\Omega_p+\delta \Omega_{pk})a_p^{\dag} a_p.
\label{eq:HBk}
\end{equation}
Equations (\ref{eq:Hhat}) and (\ref{eq:HBk}) show that in a state $|k\rangle$, the energy of an exciton $\hat{\omega}_k=\omega_k+\delta \omega_k$ is renormalized due to its coupling with the phonons. Similarly, the frequency $\hat{\Omega}_{pk}=\Omega_p+\delta \Omega_{pk}$ of phonon $p$ accompanied by an exciton in the state $|k\rangle$ is renormalized by an amount $\delta \Omega_{pk}$. The energy corrections are thus defined as
\begin{eqnarray}
\delta \omega_k &=& \sum_{p=1}^{N}\sum_{k'=1}^{N} \frac{M^{2}_{pkk'}}{\omega_k-\omega_{k'}-\Omega_p} \nonumber \\
\delta \Omega_{pk} &=& \sum_{k'=1}^{N} \frac{2 M^{2}_{pkk'} (\omega_k-\omega_{k'})}{(\omega_k-\omega_{k'})^2-\Omega_p^2}.
\label{eq:PT}
\end{eqnarray}

The operator $\hat{H}$ defines the effective exciton-phonon Hamiltonian. Being diagonal in the unperturbed basis, its eigenvalues $\epsilon_{k,n_p}=\hat{\omega}_k+\sum_{p=1}^{N} n_p \hat{\Omega}_{pk}$ stand for the system eigenenergies up to second order in $\Delta H$. 
The corresponding eigenstates are defined as $|\Psi_{k,n_p} \rangle = U^{\dag} |k, n_p \rangle$. Consequently, $U$ provides a new point of view in which $\hat{H}$ no longer describe independent excitations but refers to entangled exciton-phonon states. The entanglement results from a dual dressing effect because a state $|k\rangle$ defines an exciton dressed by a virtual phonon cloud whereas a number state $| n_p \rangle$ describes phonons clothed by virtual excitonic transitions. The behavior of the energy corrections has been studied in great details \cite{kn:pouthier5,kn:pouthier6,kn:pouthier7}. Owing to the dressing, we have shown that the excitonic eigenenergies are redshifted and the energy correction of a stationary wave $|k\rangle$ scales as 
\begin{equation}
\delta \omega_{k} \approx -E_B(1-2/L)-\frac{16\Phi E_B}{3 \pi \Omega_c} \cos(\frac{k\pi}{L}).
\label{eq:domk}
\end{equation}
By contrast, the phonon frequencies are either redshifted or blueshifted, depending on the state occupied by the exciton that accompanies the phonons. Such a mechanism is well-accounted by the following law
\begin{eqnarray}
\delta \Omega_{pk} &\approx& -\frac{16\Phi E_B}{L\Omega_c} \sin(\frac{p\pi}{2L})\cos^2(\frac{p\pi}{2L}) \nonumber \\
&\times& \left( \cos(\frac{k\pi}{L})+\frac{\delta_{p,k}-\delta{p,L-k}}{2} \right).
\label{eq:dOMpk}
\end{eqnarray}

To conclude, let us mention that for a given temperature and a fixed coupling strength, we have introduced a critical length
$L^*\approx 0.1\Omega_c^2/(E_Bk_BT)$ ($k_B$ is the Boltzmann constant) so that for $L<L^*$, PT correctly describes the exciton-phonon dynamics. By contrast, PT breaks down for $L>L^*$ owing to the occurrence of quasi-resonance between unperturbed states.

\subsection{Transport properties}

According to the standard theory of open quantum systems \cite{kn:book,kn:book1}, a complete understanding of the exciton dynamics is obtained from the knowledge of the RDM $\sigma(t)=Tr_B[\exp(-iHt)\rho(0) \exp(iHt)]$, where $Tr_B$ stands for a partial trace over the phonon degrees of freedom. The time evolution corresponds to an Heisenberg representation with respect to $H$ and the initial conditions are specified by the exciton-phonon density matrix $\rho(0)$. 

To define $\rho(0)$, let us mention that without any perturbation the lattice is assumed to be in thermal equilibrium at temperature $T$. Assuming $\omega_0 \gg k_BT$, each two-level system is in its ground state. This is no longer the case for the phonons that form a thermal bath described by the Boltzmann distribution $\rho_B=\exp(-\beta H_B)/Z_B$, where $Z_B$ is the phonon partition function ($\beta=1/k_BT$). Consequently, to study exciton-mediated energy flow, one assumes that the lattice interacts with an external source which brings the system in a state out of equilibrium. This source is resonantly coupled with the two-level systems, only, and it does not affect the remaining degrees of freedom. Moreover, it is supposed to act during a very short time scale so that the exciton is prepared adiabatically without any change in the quantum state of the phonons. To simplify the discussion, we consider a situation in which an exciton is created on the site $x_0$ at time $t=0$ so that the initial density matrix is $\rho(0)=\rho_A \otimes \rho_B$  where $\rho_A=|x_0\rangle \langle x_0|$. Note that for amide-I exciton in biopolymers, such an excitation may result from the energy released by the hydrolysis of ATP \cite{kn:helix1} or from charge neutralization upon electron capture by a protonated helix \cite{kn:helix6}. 

In that context, in the local basis, the exciton RDM is expressed as  
\begin{equation}
\sigma(x,x',t)= Tr_B[\rho_B \langle x_0|  e^{iHt} |x'\rangle \langle x| e^{-iHt}|x_0\rangle].
\label{eq:RDM}
\end{equation}
Its diagonal elements $P(x,t)=\sigma(x,x,t)$ yield the exciton density that measures the probability to observe the exciton on  site $x$ at time $t$ given that it was created on site $x_0$ at time $t=0$. The exciton density is the central object of the present study. It provides information about the way the excitonic energy flows along the lattice after its initial implementation. Moreover, its knowledge allows us to characterize the influence of the phonons on the process of energy redistribution. The off-diagonal elements of the RDM measure the ability of the exciton to develop or to maintain coherent superimpositions of local states. They provide additional information for determining whether the energy redistribution is coherent or incoherent. 

For describing the time evolution of the RDM, we do not derive a GME. Instead, we apply PT that directly provides an expression of the RDM. This procedure can be summarized as follows (see appendix B). We first introduce $U$ and diagonalize $H$ in Eq. (\ref{eq:RDM}). Second, we use the fact that $\hat{H}$ is the sum of independent contributions, each contribution describing the exciton-phonon system when the exciton occupies a specific state $|k\rangle$ (Eq.(\ref{eq:Hhat})). Then, we define the following Heisenberg representation $U_k(t)=e^{i\hat{H}_B^{(k)} t} U e^{-i\hat{H}_B^{(k)} t}$. 
Finally, we introduce the effective phonon density matrix 
\begin{equation}
\rho_B^{(kk')}(t)=\frac{\exp \left[ - \beta H_B-it(\hat{H}_B^{(k)}-\hat{H}_B^{(k')}) \right]}{Z_B^{(kk')}(t)},
\label{eq:rhoBkk'}
\end{equation}
where
\begin{equation}
Z_B^{(kk')}(t)=Tr_B e^{  - \beta H_B-it(H_B^{(k)}-H_B^{(k')})}.
\end{equation}
Strictly speaking, $\rho_B^{(kk')}(t)$ is not a density matrix because it yields complex values for the phonon population. However, it is isomorphic to $\rho_B$ with the correspondence $\beta \Omega_p \rightarrow  \beta \Omega_p+i(\delta \Omega_{pk}-\delta \Omega_{pk'})t$. It thus yields averages equivalent to thermal averages and facilitates the calculations. In that context, after simple algebraic manipulations, the RDM is rewritten as 
\begin{widetext}
\begin{equation}
\sigma (x,x',t) =\sum_{k=1}^N \sum_{k'=1}^N \frac{Z_B^{(kk')}(t)}{Z_B} e^{-i(\hat{\omega}_{k}-\hat{\omega}_{k'})t}  
Tr_B \left[ \rho_B^{(kk')}(t) \langle x_0| U^{\dag}_{k'}(-t) |k'\rangle \langle k' | U |x'\rangle  \langle x|U^{\dag}|k\rangle \langle k | U_{k}(-t)|x_0\rangle \right].
\label{eq:RDM2} 
\end{equation}
\end{widetext}
The final step, quite fastidious, consists in expanding $U$ as a Taylor series with respect to $\Delta H$. By bringing together the various terms, one finally obtains the second order expression of $\sigma(x,x',t)$ that will be used in the following of the text. Note that we felt it was unnecessary to give the expression of the RDM that exhibits about thirty different contributions. 

To conclude, let us mention that in the weak-coupling limit, the exciton-phonon coupling mainly induces a renormalization of the system energies without significantly modifying the quantum states. The exciton-phonon eigenstates basically correspond to the unperturbed states so that the transformation $U$ behaves as the unit operator in Eq.(\ref{eq:RDM2}). One thus obtains the so-called zero order approximation of the RDM written as
\begin{eqnarray}
\sigma_0 (x,x',t) &=& \sum_{k=1}^N \sum_{k'=1}^N \frac{Z_B^{(kk')}(t)}{Z_B} e^{-i(\hat{\omega}_{k}-\hat{\omega}_{k'})t} \nonumber \\ 
& \times &\langle x_0| k'\rangle \langle k' | x'\rangle  \langle x | k\rangle \langle k | x_0\rangle.  
\label{eq:RDM3} 
\end{eqnarray} 

\section{Numerical results}

In this section, Eq.(\ref{eq:RDM2}) is used for studying the excitonic energy flow in a finite-size lattice. Despite its general nature, the previous formalism will be applied for investigating energy redistribution in a lattice of H-bonded peptide units, a system for which the parameters are well-known \cite{kn:helix1,kn:helix2,kn:helix3,kn:edler,kn:helix4,kn:helix5,kn:helix6,kn:helix7,kn:helix8} (Note that this formalism can be used for describing many situations in which a narrow-band exciton interacts with the phonons of the lattice such as Frenkel exciton in molecular aggregates \cite{kn:may} or vibrational exciton in adsorbed nanostructures \cite{kn:surf1,kn:surf2,kn:surf3,kn:surf4}). In such a lattice, peptide units linked by H bonds are regularly distributed. Each site contains an amide-I mode (C=O vibration) that gives rise to a vibrational exciton. This exciton interacts with the H bond vibrations that form a bath of acoustic phonons at temperature $T$. 
To study the amide-I dynamics, the following parameters are used: $\omega_0=1660$ cm$^{-1}$, $W=15$ Nm$^{-1}$, $M=1.8\times 10^{-25}$ kg, $\Omega_c=96.86$ cm$^{-1}$ and $\Phi=7.8$ cm$^{-1}$. To avoid PT breakdown, the coupling strength will vary around $\chi=10$ pN \cite{kn:helix8}.
Of course, the model displayed in Sec. II is too simple to accurately describe vibrational energy flow in a real biopolymer whose dynamics exhibits a tremendous complexity due to the large number of degrees of freedom. In particular, this model is unable to account for the finite amide-I lifetime because it conserves the number of amide-I exciton. Nevertheless, it involves ingredients that play a key role in interpreting specific experiments such as pump-probe spectroscopy in $\alpha$-helices \cite{kn:edler} and Electron Capture Dissociation in finite-size polypeptides\cite{kn:helix6}. Consequently, we do not claim that the model is relevant to explain in details the vibrational dynamics in a protein. Nevertheless, its interest lies in the fact that it provides a simple approach to promote the idea that the confinement modifies the exciton energy redistribution. 

\begin{figure}
\includegraphics{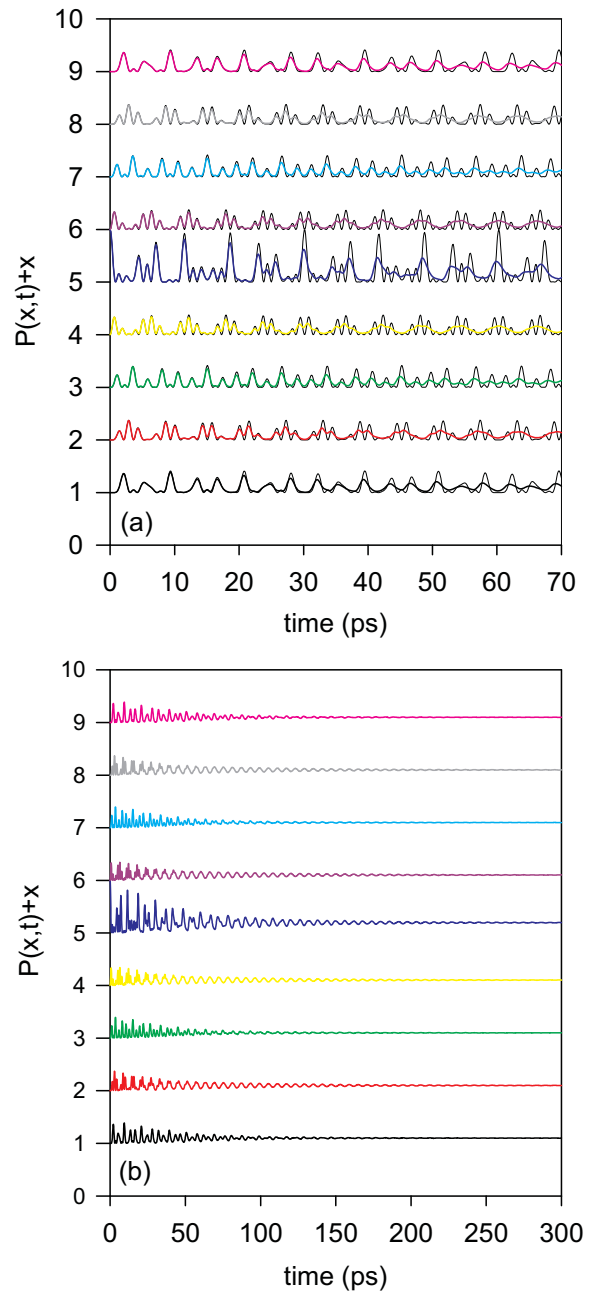}
\caption{Time evolution of the exciton density $P(x,t)$ for $N=9$, $T=300$ K, $\chi=10$ pN and $x_0=5$. (a) Short-time evolution and (b) time evolution over a longer time scale. Note that thick lines in Fig. 1a refer to a coherent energy transfer that arises when the relaxation induced by the phonons is neglected (see the Sec. IV).}
\end{figure}

\subsection{Time evolution of the exciton density}

As illustrated in Fig. 1 for $N=9$, $T=300$ K, $\chi=10$ pN and $x_0=5$, the evolution of the exciton density is governed by several time scales whose duration depends on the model parameters. 
In the short-time limit ($t<20-30$ ps), Fig. 1a shows that $P(x,t)$ flows almost coherently along the lattice. Note that a purely coherent motion is illustrated by the thick lines in Fig. 1a, as discussed in the Sec. IV. The exciton behaves as a confined wave that experiences reflections on the lattice sides. Indeed, after the initial excitation, two wave packets propagate on each side of the central site $x_0$. At $t\approx 0.4$ ps, the wave packets have left the excited region and more than $50$\% of the initial energy occur on the neigboring sites $x_0\pm1$. At $t\approx 2.1$ ps, the wave packets reach the lattice sides where the exciton density reaches $0.36$. Then, the wave packets are reflected and they propagate back to the central site. Consequently, $71$\% of the initial energy recur on the central site at $t\approx 7.1$ ps. Such a mechanism takes place almost periodically so that quantum recurrences occur. Therefore, $P(x_0,t)$ shows a series of peaks. The first peaks take place at $t=11.5$ ps, $t=18.6$ ps and $t=23.0$ ps, the corresponding amplitudes being equal to $0.81$, $0.74$ and  $0.53$.

Over a longer time scale, the coherent nature of the exciton tends to disappear (Fig. 1b). The recurrences in $P(x_0,t)$ become less and less pronounced. For instance, when the exciton moves coherently, $P(x_0,t)$ exhibits two strong recurrences at $t=47.4$ ps and $t=105.9$ ps, the corresponding amplitudes being $0.98$ and $0.97$ (the second recurrence is not drawn). By contrast, when the exciton-phonon coupling is taken into account, $P(x_0,t)$ remains smaller than $0.46$ in the neighborhood of the first recurrence and it does not exceed $0.27$ in the neighborhood of the second recurrence. In fact, $P(x,t)$ exhibits several spectral components. It thus develops a rather complex time evolution quite similar to that of an almost periodic function, the almost period being the signature of the recurrences. However, as time increases, specific spectral components disappear so that, after approximately $100$ ps, $P(x,t)$ reduces to a damped sine function (Fig. 1b). These features clearly characterize a relaxation mechanism that affects the energy redistribution over an intermediate time scale.  

Finally, in the long-time limit, all the spectral components vanish. The density converges toward a stationary distribution that specifies a new equilibrium density $P_{eq}(x)$. With the parameters used in Fig.1, $P_{eq}(x)$ is not uniformly distributed. At $t=500$ ps, we obtain $P(x_0,t)\approx 0.2$ whereas 
$P(x,t)\approx 0.1$, $\forall x\neq x_0$, suggesting that the exciton keeps the memory of its initial state despite its interaction with the phonons. Note that for $\chi=10$ pN, we have verified that the zero order approximation of the RDM (Eq.(\ref{eq:RDM3})) yields a quite good estimate of the time evolution of the density. 

\begin{figure}
\includegraphics{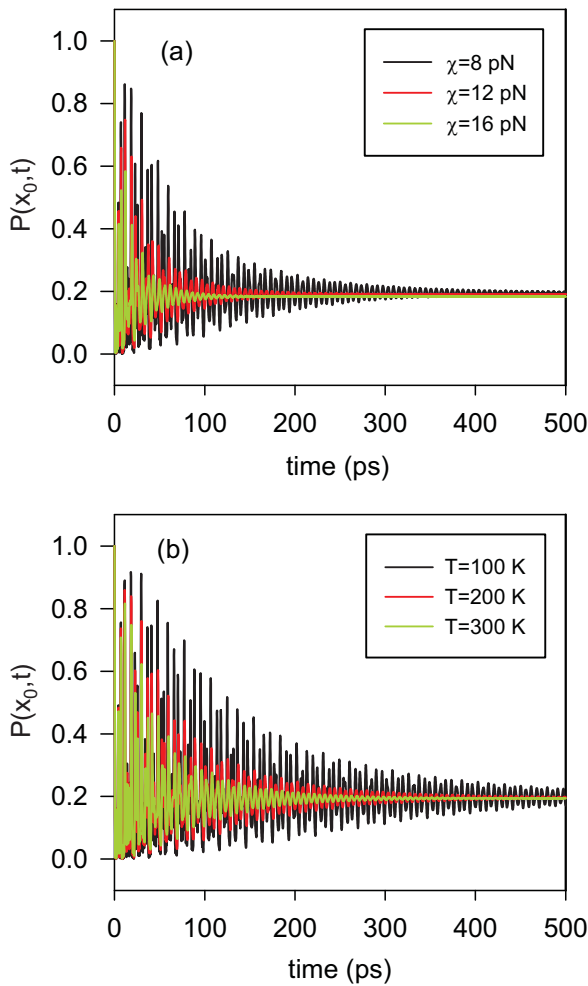}
\caption{Time evolution of the exciton density $P(x_0,t)$ for $N=9$ and $x_0=5$. (a) Influence of the exciton-phonon coupling strength for $T=300$ K and (b) influence of the temperature for $\chi=10$ pN.}
\end{figure}

The rate at which the exciton reaches the equilibrium depends on both the coupling strength and the temperature. This feature is illustrated in Fig.2 that displays the time evolution of the density on the excited site $P(x_0,t)$ for different parameter values. As shown in Fig. 2a, the coupling $\chi$ enhances the relaxation rate. The larger $\chi$ is, the faster the $P_{eq}(x_0)$ is reached. To estimate the relaxation time $\tau_R$, we assume that $P(x_0,t)$ scales as a rapidly varying signal dressed by a smooth envelope function that describes an exponential decay toward equilibrium. In doing so, one successively obtains $\tau_R\approx100$ ps, $\tau_R\approx40$ ps and $\tau_R\approx20$ ps for $\chi=8$ pN, $\chi=12$ pN and $\chi=16$ pN, indicating that $\tau_R\propto1/\chi^2$. Similarly, as displayed in Fig. 2b, the temperature also enhances the relaxation rate. One thus obtains $\tau_R\approx190$ ps, $\tau_R\approx90$ ps and $\tau_R\approx60$ ps for $T=100$ pN, $T=200$ pN and $T=300$ pN. Such a behavior suggests that the relaxation time is inversely proportional to $T$. Note that the zero order RDM (Eq.(\ref{eq:RDM3})) yields results almost identical to those displayed in Fig. 2.

\begin{figure}
\includegraphics{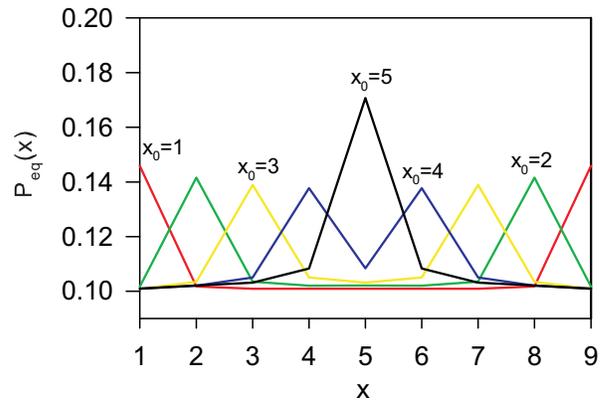}
\caption{Influence of the initial position of the exciton $x_0$ on the equilibrium distribution $P_{eq}(x)$ for $N=9$, $\chi=10$ pN and $T=300$ K.}
\end{figure}

\subsection{Equilibrium distribution} 

The influence of the initial position of the exciton on the equilibrium distribution is illustrated in Fig. 3 for $N=9$, $\chi=10$ pN and $T=300$ K. Although $P_{eq}(x)$ is invariant under mirror symmetry, that is,  $P_{eq}(x)=P_{eq}(L-x)$, two distinct situations occur. 
When $x_0=L/2$, $P_{eq}(x)$ is partially localized on the excited site. It is almost uniformly distributed over the sites $x\neq x_0$ ($P_{eq}(x)\approx 0.1$ $\forall x \neq x_0$) whereas it is approximately two times larger on the excited site ($P_{eq}(x_0)\approx 0.17$). In fact, when $\chi$ nearly vanishes or at very low temperature, $P_{eq}(x)$ exhibits an universal behavior. It depends only on the lattice size and its scales as $P_{eq}(x)\approx(1+\delta_{xx_0})/L$.
By contrast, when $x_0$ is not located at the center of the lattice, $P_{eq}(x)$ partially localizes over the two sites $x_0$ and $L-x_0$. It remains almost uniform over the sites $x\neq x_0$ and $x\neq L-x_0$ ($P_{eq}(x)\approx 0.1$). However, it becomes approximately 1.5 times larger on the sites $x_0$ and $L-x_0$. One thus obtains $P_{eq}(x_0)=P_{eq}(L-x_0)=0.146$, $0.142$, $0.139$ and $0.138$ for $x_0=1$, $2$, $3$ and $4$, respectively. Note that when $\chi$ nearly vanishes or at very low temperature, we have observed the following universal behavior: $P_{eq}(x_0)=P_{eq}(L-x_0)\approx3/2L$ and $P_{eq}(x)\approx1/L$ for $x\neq x_0$ and $x\neq L-x_0$.

\begin{figure}
\includegraphics{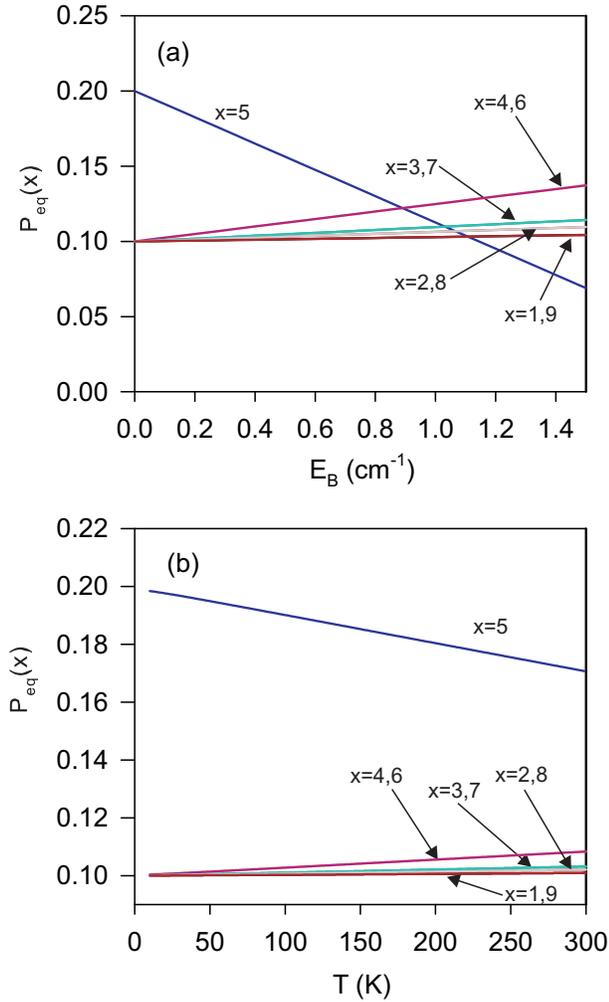}
\caption{Parameter dependence of the equilibrium distribution for $N=9$ and $x_0=5$. (a) Influence of the exciton-phonon coupling for $T=300$ K and (b) influence of the temperature for $\chi=10$ pN.}
\end{figure}

The equilibrium distribution is also quite sensitive to the coupling and to the temperature, as shown in Fig. 4 for $N=9$ and $x_0=L/2$. 
When $\chi$ increases (Fig. 4a), $P_{eq}(x_0)$ decreases whereas $P_{eq}(x)$ increases $\forall x\neq x_0$. Whatever the $x$ values, $P_{eq}(x)$ scales linearly with $E_B$. Therefore, when $\chi\approx 16$ pN ($E_B\approx 0.86$ cm$^{-1}$), the equilibrium density becomes almost uniform in the neighborhood of the excited site. One thus obtains $P_{eq}(x_0)\approx P_{eq}(x_0\pm1)\approx0.12$. Then, when $\chi \approx 18$ pN ($E_B\approx 1.08$ cm$^{-1}$) a hole occurs in the exciton density at the center of the lattice, that is, $P_{eq}(x_0)<P_{eq}(x)$ $\forall x\neq x_0$. This hole continues to deepen when $\chi$ increases again. Note that, as discussed in Sec. II.B, PT is valid provided that $\chi$ remains smaller than a critical value. With the parameters used in Fig. 4a, this critical value is approximately $\chi\approx20$ pN ($E_B\approx 1.4$ cm$^{-1}$). Above this value, unphysical results were obtained, that is, $P_{eq}(x_0)<0$ when $\chi>26$ pN.
As shown in Fig. 4b, the influence of the temperature is quite similar to that of the coupling. Indeed, as $T$ increases,  $P_{eq}(x_0)$ decreases whereas $P_{eq}(x)$ increases $\forall x\neq x_0$. Note that the density evolves linearly with the temperature. The key point is that over a broad temperature range, the temperature slightly affects the equilibrium distribution. For instance, $P_{eq}(x_0)$ decreases from $0.2$ to $0.17$ when the temperature increases from $0$ to $300$ K. Consequently, the exciton density does not become uniform in the neighborhood of the excited site and it does not exhibit a hole at the center of the lattice. 
Finally, a numerical fit of the data has revealed that $P_{eq}(x_0)\approx2/L-3\eta(B) E_Bk_BT/\Omega_c^2$, where $\eta(B)$ is a slowly varying function of the adiabaticity $B=2\Phi/\Omega_c$ quite close to unity. 

\begin{figure}
\includegraphics{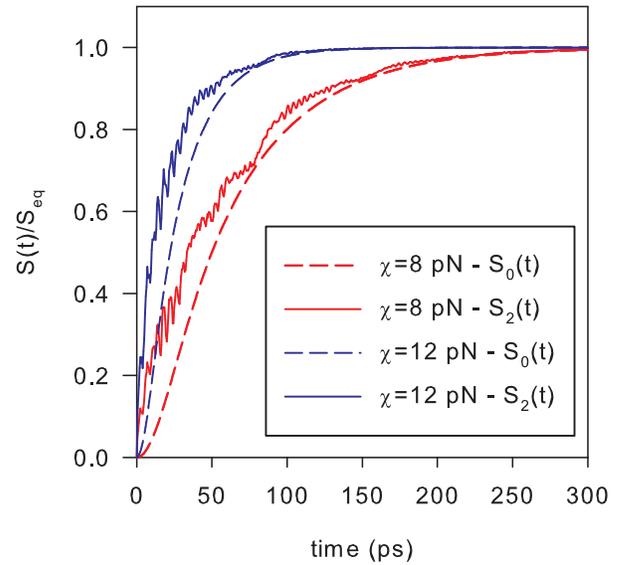}
\caption{Time evolution of the linear entropy for $N=9$, $x_0=5$ and $T=300$ K.  The entropy $S_0(t)$ (dashed lines) is defined in terms of the RDM $\sigma_0(t)$ (Eq.(\ref{eq:RDM3})) whereas the entropy $S_2(t)$ (solid lines) is obtained from the RDM $\sigma(t)$ (Eq.(\ref{eq:RDM2})).}
\end{figure}

\subsection{Linear Entropy and relaxation rate}

Let us now focus our attention on the relaxation process. As shown in Figs. 1 and  2, it is not easy to define a relaxation time because the exciton density exhibits several spectral components that decay according to different relaxation rates. To overcome this difficulty, a single relaxation time must be extract from a more "global observable". To proceed, we consider the linear entropy defined as $S(t)=1-Tr_A[\sigma(t)^2]$. It provides a measure of the missing information about the state of the exciton when compared with a situation in which the exciton evolves freely. It accounts for the arrow of time and it describes the relaxation mechanism as a consequence of an information transfer "from the exciton toward the phonon bath" in the form of exciton-phonon correlations \cite{kn:book}. 

The behavior of the entropy is shown in Fig. 5 for $N=9$, $x_0=5$, $T=300$ K and for two $\chi$ values. Two kinds of calculations have been carried out by introducing the zero order entropy $S_0(t)$ (dashed lines), defined in terms of the RDM $\sigma_0(t)$ (Eq.(\ref{eq:RDM3})), and the second order entropy $S_2(t)$ (solid lines), defined in terms of the RDM $\sigma(t)$ (Eq.(\ref{eq:RDM2})). The entropy $S_0(t)$ is a monotonic function that increases with time. Initially equal to zero, it first scales as $t^2$ in the very short-time limit. Then, it increases as time increases by following an exponential function that rises to a maximum. In the long-time limit, it converges to a constant value $S_{\infty}$, quite close to unity, that depends only on $L$ and $x_0$.
When the entangled nature of the exciton-phonon eigenstates is taken into account, the time evolution of the linear entropy slightly changes. As previously, $S_2(t)$ increases from zero to reach a constant value. In the intermediate-time limit and in the long-time limit, it still behaves as an exponential function that rises to a maximum. However, two main differences occur. First, in the very short-time limit, $S_2(t)$ rapidly increases over a few tenths of ps. Then, the entropy no longer defines a monotonic function. Instead, it behaves as an increasing function that supports high-frequency small-amplitude oscillations. 

Such a behavior allows us to introduce a phenomenological measure of the relaxation rate. To proceed, we works with the zero order entropy whose evaluation is must easier. We thus assume that it scales as $S_0(t)\approx S_{\infty}(1-\exp(-2\Gamma_R t))$, the factor $2$ in the exponential accounting for the fact that $S_0(t)$ involves the square of the RDM. Of course, this expression works quite well in the intermediate-time limit and in the long-time limit, but it fails in reproducing the short-time behavior of the entropy. So defined, the parameter $\Gamma_R$ represents an effective relaxation rate that provides information on the time $\tau_R=1/\Gamma_R$ needed to the exciton to reach the equilibrium. Note that, in the weak-coupling limit, it turns out that there is no much difference between the relaxation rates provided by the two entropies.
\begin{figure}
\includegraphics{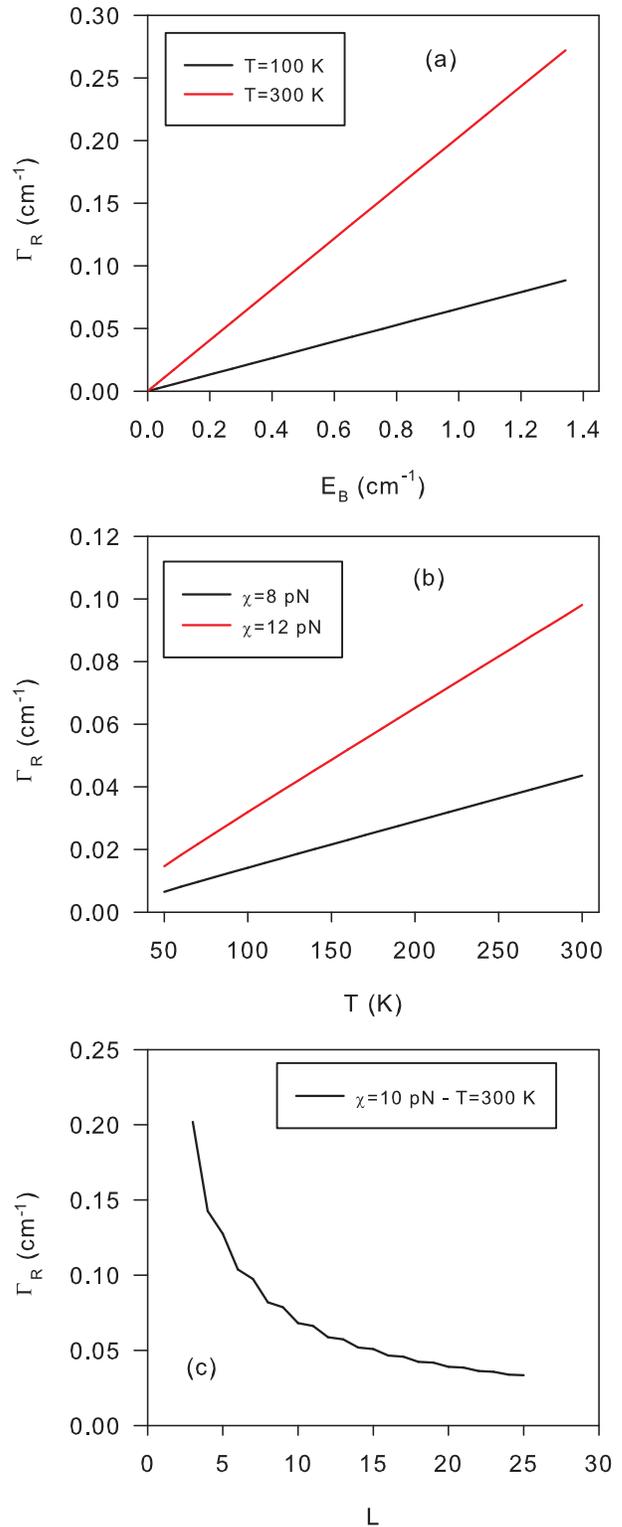}
\caption{Behavior of the relaxation rate $\Gamma_R$. (a) Influence of the exciton-phonon coupling for $N=9$, $x_0=1$ and for two values of the temperature. (b) Influence of the temperature for $N=9$, $x_0=1$ and for two values of the coupling. (c) Influence of the lattice size for $x_0=1$, $\chi=10$ pN and $T=300$ K.}
\end{figure}

The behavior of the relaxation rate $\Gamma_R$ is illustrated in Fig. 6 for $x_0=1$. As shown in Fig. 6a, $\Gamma_R$ increases linearly with $E_B$. For instance, at $T=300$ K, $\Gamma_R$ ranges from
$0.068$ cm$^{-1}$ ($\tau_R=77.85$ ps) to $0.153$ cm$^{-1}$ ($\tau_R=34.59$ ps) when $\chi$ varies from 10 pN ($E_B=0.33$ cm$^{-1}$) to 15 pN ($E_B=0.75$ cm$^{-1}$). Similarly, Fig. 6b reveals that $\Gamma_R$ increases almost linearly with the temperature over a broad temperature range. For instance, for $\chi=12$ pN ($E_B=0.48$ cm$^{-1}$), $\Gamma_R$ extends from
$0.015$ cm$^{-1}$ ($\tau_R=361.90$ ps) to $0.098$ cm$^{-1}$ ($\tau_R=54.10$ ps) when $T$ varies from 50 K to 300 K. Note that we have also verified that $\Gamma_R$ increases almost linearly with the adiabaticity $B$ provided that the nonadiabatic limit is reached. Similarly, $\Gamma_R$ slightly depends on $x_0$. A rather fast relaxation occurs for $x_0=L/2$ whereas a quite slow relaxation takes place when $x_0=2$ or $x_0=L-2$. For instance, for $N=9$, $\chi=10$ pN and $T=300$ K, we obtained $\Gamma_R=0.087$ cm$^{-1}$ ($\tau_R=60.50$ ps) for $x_0=L/2$ and $\Gamma_R=0.065$ cm$^{-1}$ ($\tau_R=81.50$ ps) for $x_0=2$.
Finally,  the most surprising effect arises from the dependence of the rate with respect to the lattice size $L$. Indeed, as illustrated in Fig. 6c, $\Gamma_R$ decreases with $L$. It approximately scales as $\Gamma_R\propto 1/L$ so that, for $\chi=10$ pN and $T=300$ K, it decreases from $0.068$ cm$^{-1}$ ($\tau_R=77.85$ ps) to $0.039$ cm$^{-1}$ ($\tau_R=135.97$ ps) when $L$ varies from 10 to 20. 
This result suggests that the relaxation mechanism is an intrinsic property of the finite-size lattice. The relaxation is thus enhanced by the confinement so that the shorter the lattice is, the larger is the relaxation rate . 
To conclude, let us mention that the previous data were exploited to extract an analytical expression of the relaxation rate. Therefore, in the nonadiabatic weak-coupling limit, it turns out that the rate scales as 
\begin{equation}
\Gamma_R\approx \alpha(x_0) \frac{ BE_Bk_BT}{\Omega_cL},
\label{eq:gammaR1}  
\end{equation}
where $\alpha(x_0)\approx 6.5\pm1$ slightly depends on the initial position of the exciton. 

\section{Discussion}

As shown in the previous section, our numerical results contrast sharply with what happens in an infinite lattice with translational invariance. Indeed, in this latter situation, spatial correlations in the phonon bath prevent the exciton diffusivity. The exciton develops a wavelike motion so that it propagates coherently along the lattice as if it was insensitive to the phonons. In a confined lattice, a fully different behavior takes place. Although the exciton delocalizes coherently in the short-time limit, its interaction with the phonons favors a relaxation mechanism when time elapses. Consequently, the coherent nature of the exciton density gradually disappears. The density finally converges toward an almost uniform  equilibrium distribution that slightly keeps the memory of the initial exciton position. Of course, in agreement with what one expects from a physical point of view, the relaxation rate is enhanced by the coupling strength and by the temperature. However, more surprisingly, we have observed that an increase of the lattice size softens the influence of the phonon bath resulting in a slowdown in the relaxation process. In other words, the relaxation previously evidenced is an intrinsic property of a confined lattice.

\subsection{Approximate expression of the exciton density}

To understand the numerical results, let us discuss the physics which derives from the expression of the exciton density. According to Eq.(\ref{eq:RDM}) and by expanding the partial trace in the phonon number state basis, the density is rewritten as 
\begin{equation}
P(x,t)=\sum_{n_p,m_p} \langle n_p|\rho_B|n_p\rangle \left| \langle m_p, x | e^{-iHt} |x_0,n_p\rangle \right|^2.
\label{eq:PX1}
\end{equation}
So defined, $P(x,t)$ characterizes a generalized quantum probability. It measures the probability to observe the exciton-phonon system in the factorized state $|x,m_p\rangle$  at time $t$ provided that it was in the state $|x_0,n_p\rangle$  at time $t=0$. Because we are concerned with the exciton dynamics, a sum over all possible phonon states at time $t$ is performed and an average over the initial phonon state is realized. 

To evaluate $P(x,t)$, one can take advantage of the fact that in the nonadiabatic weak-coupling limit, the exciton-phonon coupling mainly induces a renormalization of the system energies without significantly modifying the quantum states. As a result, the Hamiltonian $H$ in Eq.(\ref{eq:PX1}) can be replaced by the effective Hamiltonian $\hat{H}$ (Eq.(\ref{eq:Hhat})) so that the density becomes
\begin{equation}
P(x,t)=\left< \left| \sum_{k=1}^N  e^{-i\hat{\omega}_k t}  e^{-i\sum_p n_p\hat{\Omega}_{pk}t} \langle x |k \rangle\langle k|x_0\rangle\right|^2 \right>
\label{eq:PX2}
\end{equation}
where the symbol $<...>$ stands for an average over the initial phonon number state. 
In accordance with the laws of quantum mechanics, before the average, the exciton density reduces to the square modulus of the probability amplitude that the exciton tunnels from $x_0$ to $x$ during time $t$, the phonons evolving freely from their initial state $|n_p\rangle$. This amplitude is the sum over the elementary probability amplitudes associated to the different paths that the exciton can follow to tunnel. A given path defines a transition through a stationary wave $|k\rangle$ and it exhibits three contributions. First, it involves  the weight of the localized states $|x\rangle$ and $|x_0\rangle$ in the stationary wave $|k\rangle$. Then, it depends on a phase factor that accounts for the free evolution of the exciton. Finally, it involves a second phase factor that refers to the quantum evolution of the phonons dressed by the exciton in the state $|k\rangle$. In that context, by performing the average over the phonon degrees of freedom, the exciton density is finally expressed as 
\begin{eqnarray}
P (x,t) &\approx & \sum_{k=1}^N \sum_{k'=1}^N F(kk',t) e^{-i(\hat{\omega}_{k}-\hat{\omega}_{k'})t} \nonumber \\ 
& \times &\langle x_0| k'\rangle \langle k' | x'\rangle  \langle x | k\rangle \langle k | x_0\rangle, 
\label{eq:PX3} 
\end{eqnarray} 
where $F(kk',t)$ stands for the so-called decoherence factor defined as 
\begin{equation}
F(kk',t)=\left< e^{-i\sum_p n_p(\hat{\Omega}_{pk}-\hat{\Omega}_{pk'})t} \right>\equiv \frac{Z_B^{(kk')}(t)}{Z_B}.
\label{eq:FACTOR}
\end{equation}
At this step, it is straightforward to show that Eq.(\ref{eq:PX3}) corresponds to the diagonal element of the zero order RDM $\sigma_0(t)$ displayed in Eq.(\ref{eq:RDM3}). Provided that $k\neq k'$, $F(kk',t)$ defines a decaying function that tends to zero in the long-time limit \cite{kn:pouthier6}. It represents the way the phonon bath destroys the coherence of an excitonic state defined as the superimposition between two stationary waves $|k\rangle$ and $|k'\rangle$. Therefore $F(kk',t)$ accounts for quantum decoherence and it is expressed as 
\begin{equation}
F(kk',t)=\prod_{p=1}^{N} \frac{1-e^{-\beta \Omega_p}}{1-e^{-\beta \Omega_p-i(\delta \Omega_{pk}-\delta \Omega_{pk'}) t}}.
\label{eq:FACTOR1}
\end{equation}

According to Eq.(\ref{eq:PX3}), the time evolution of the exciton density $P(x,t)$ results from the quantum interferences between the different paths that the exciton can follow to tunnel from $x_0$ to $x$. Owing to these interferences, $P(x,t)$ is expressed as the sum of different spectral components, a specific component being associated to a phase factor involving an excitonic Bohr frequency. In addition, each component is modulated by a decoherence factor that results form the fact that the phonons evolve differently depending on the path followed by the exciton. More precisely, when the exciton follows two different paths, the phonons develop two different quantum evolutions so that an additional phase factor occurs. Strictly speaking, this phase factor is not responsible for decoherence. However, because the phonons are initially described by a statistical mixture of number states, an average is required. This average yields a sum over phase factors that interfere with the other giving rise to quantum decoherence. 

As shown previously \cite{kn:pouthier6}, the decoherence factor behaves as a gaussian function whose time evolution can be extracted from the short-time limit of Eq.(\ref{eq:FACTOR1}). One finally obtains  
\begin{equation}
F(kk',t)\approx e^{-i\sum_p \bar{n}_p(\hat{\Omega}_{pk}-\hat{\Omega}_{pk'})t}e^{-\Gamma_{kk'}^2 t^2},
\label{eq:FACTOR2}
\end{equation}
where $\bar{n}_p=[\exp(\beta \Omega_p)-1]^{-1}$ is the average number of phonons of the $p$th mode. The so-called decoherence rate $\Gamma_{kk'}$, inversely proportional to the corresponding relaxation time $\tau_{kk'}=1/\Gamma_{kk'}$, is defined as
\begin{equation}
\Gamma_{kk'}=\sqrt{\frac{1}{2} \sum_{p=1}^N \Delta \bar{n}_p^2 (\hat{\Omega}_{pk}-\hat{\Omega}_{pk'})^2},
\label{eq:GAMMA}
\end{equation}
where $\Delta \bar{n}_p^2=\bar{n}_p(\bar{n}_p+1)$ accounts for the thermal fluctuations of the phonon number.

\subsection{Coherent dynamics in the short-time limit}

By combining Eqs.(\ref{eq:PX3}) and (\ref{eq:FACTOR2}), we are able to formally explain the behavior of the exciton density as follows. 
In the short-time limit, that is, over a time scale shorter than the smallest relaxation time, each decoherence factor reduces to  an undamped phase factor so that the density is approximately expressed as  
\begin{equation}
P(x,t)\approx \left| \sum_{k=1}^N e^{-i\tilde{\omega}_k t} \langle x |k \rangle \langle k|x_0\rangle \right|^2, 
\label{eq:PX4}
\end{equation}
where $\tilde{\omega}_k$ defines the effective energy of the exciton in state $|k\rangle$, as
\begin{equation}
\tilde{\omega}_k=\omega_k+\delta \omega_k + \sum_{p=1}^N \bar{n}_p \delta \Omega_{pk}.
\label{eq:effom}
\end{equation}
As illustrated by the solid lines in Fig. 1a, Eq.(\ref{eq:PX4}) shows that the exciton moves freely as if it was insensitive to the phonon bath. A coherent energy transfer takes place, this transfer being characterized by quantum recurrences that result from the confinement of the lattice. Nevertheless, although they preserve the coherent nature of the exciton propagation, the phonons affect the dynamical parameters that govern this wavelike motion through the concept of effective energy. Indeed, as shown in Eq.(\ref{eq:effom}), it is as if the exciton was characterized by temperature dependent eigenvalues $\tilde{\omega}_k$ that exhibit two corrections. The first correction, encoded in the parameters $\delta \omega_k$, accounts for the fact that during its propagation the exciton is dressed by a virtual cloud of phonons. The second correction, $\sum_p \bar{n}_p \delta \Omega_{pk}$, results form the phase of the decoherence factors. 

In that context, a moment's reflection will convince the reader that $\tilde{\omega}_k$ reduces to the small polaron energy in a state $|k\rangle$ obtained by using a finite temperature mean field theory (see for instance Ref. \cite{kn:trans4}). Within this theory, the exciton-phonon coupling is partially removed by performing the so-called Lang-Firsov transformation. This transformation provides a new point of view in which the elementary excitation is no longer an exciton but a small polaron, that is, a composite particle that describes an exciton dressed by a lattice distortion. Then, a mean field approach is invoked to treat the remaining polaron-phonon interaction, i.e. an average is realized over the phonon degrees of freedom according to the phonon density matrix $\rho_B$. In doing so, it turns out that the polaron energy in a state $|k\rangle$ is expressed as 
\begin{equation}
\omega_{k}^{(pol)}=\omega_0-\epsilon_B+2\Phi e^{-s(T)} \cos(\frac{k \pi}{L}),
\end{equation}
where $\epsilon_B=E_B(1-2/L)$ generalizes the concept of small polaron binding energy in a confined lattice \cite{kn:pouthier7} and whereas $s(T)$ is the well-known temperature dependent band-narrowing factor. The band-narrowing factor reduces to $s(0)=8E_B/3\pi\Omega_c$ at zero temperature where it scales as $s(T)=4E_Bk_BT/\Omega_c^2$ at high temperature. In that context, by inserting Eqs.(\ref{eq:domk}) and (\ref{eq:dOMpk}) into Eq.(\ref{eq:effom}), it is straightforward to show that the effective energy $\tilde{\omega}_k$ reduces to the polaron energy  $\omega_{k}^{(pol)}$ in the weak-coupling limit, that is, for $s(T)\ll1$. This result is quite interesting because it establishes the link between the present approach, based on a perturbative treatment of the exciton-phonon entanglement, and the mean field procedure usually introduced in the small polaron theory. Within PT, the temperature dependence of the effective energy originates in the modification of the quantum evolution of the phonons owing to exciton-phonon correlations.

\subsection{Relaxation mechanism}

Over a time scale longer than the shortest relaxation time but shorter than the longest relaxation time, the decaying behavior of the decoherence factors can no longer be disregarded. Therefore, a relaxation mechanism takes place resulting in the decay of the different spectral components of the exciton density. Each component disappears according to a specific decoherence rate.
As shown in Eq.(\ref{eq:GAMMA}), the temperature dependence of $\Gamma_{kk'}$ is encoded in the fluctuations of the phonon numbers. Because $\Delta \bar{n}_p$ approximately scales as $ k_BT/\Omega_p$, $\Gamma_{kk'}$ increases linearly with the temperature. 
Note that this dependence differs from the standard expression of the relaxation rate that characterizes the dephasing of an open system coupled with a reservoir of harmonic oscillators \cite{kn:may}. Indeed, in most situations the temperature dependence of the rate originates in its dependence with respect to the average phonon number $\bar{n}_p$. At high temperature, both approaches yield a similar temperature dependence since $\Delta \bar{n}_p \approx \bar{n}_p \approx k_BT/\Omega_p$. This is no longer the case at low temperature since $\Delta \bar{n}_p/\bar{n}_p=\exp(\beta \Omega_p/2)$. 
Moreover, $\Gamma_{kk'}$ is proportional to $E_B$ through its dependence with respect to the phonon energy corrections (see Eq.(\ref{eq:dOMpk})). From these dependences, we thus recover that both the temperature and the coupling strength enhance the relaxation mechanism, as observed in Fig. 2. In that context, it is obvious that the duration of the relaxation mechanism will be approximately given by the longest relaxation time $\tau_R=Max[\tau_{kk'}]$, that is, the invert of the smallest decoherence rate $\Gamma_R=Min[\Gamma_{kk'}]$. From Eq.(\ref{eq:GAMMA}), it turns out that the smallest decoherence rate is $\Gamma_{12}$. Therefore, by inserting Eq.(\ref{eq:dOMpk}) into Eq.(\ref{eq:GAMMA}), one obtains a typical expression for the relaxation rate, as
\begin{equation}
\Gamma_R=\frac{4BE_B k_BT}{\Omega_c L}\sqrt{1+\frac{27\pi^4}{16L^3}}.
\label{eq:gammaR2}
\end{equation}
Equation (\ref{eq:gammaR2}), quite similar to the expression of the rate extracted form numerical data (see Eq.(\ref{eq:gammaR1})), 
clearly evidences the role of both the exciton-phonon coupling and the temperature. In addition, it shows that the relaxation is an intrinsic property of a confined lattice, the rate $\Gamma_R$ decreasing with the lattice size $L$. Consequently, the larger is the lattice size, the smaller is the relaxation rate. Such an effect originates in the size dependence of the energy correction of the phonons that scales as $1/L$, as shown in Eq.(\ref{eq:dOMpk}). 

\subsection{Equilibrium distribution}

Finally, in the long-time limit, that is, over a time scale longer than the longest relaxation time, all the spectral components of $P(x,t)$ have disappeared owing to quantum decoherence. The exciton density converges toward a stationary distribution defined as 
\begin{equation}
P_{eq}(x)\approx \sum_{k=1}^N \left|  \langle x |k \rangle \langle k|x_0\rangle \right|^2. 
\label{eq:PX5}
\end{equation}
Equation (\ref{eq:PX5}) represents a time independent "classical" probability, the word classical meaning that there is no longer quantum interferences. It thus reduces to the sum of the probabilities associated to the realization of each path that the exciton can follow to tunnel from $x_0$ to $x$. Consequently, $P_{eq}(x)$ no longer depends on the temperature nor on the coupling strength. It involves only the lattice size and the initial position of the exciton. It is as if the exciton kept the memory of its initial location so that, for $x_0=L/2$, the equilibrium distribution satisfies  
\begin{equation}
P_{eq}(x)= \left\{ 
\begin{array}{ll}
2/L & \mbox{if $x=x_0$}  \\
1/L & \mbox{if $x\neq x_0$}.
\end{array}
\right.
\end{equation}
Otherwise, for $x_0\neq L/2$, the equilibrium distribution is defined as 
\begin{equation}
P_{eq}(x)= \left\{ 
\begin{array}{ll}
1.5/L & \mbox{if $x=x_0$ and $x=L-x_0$} \\
1/L & \mbox{otherwise}.
\end{array}
\right.
\end{equation}
In a quite good agreement with the results displayed in Fig. 3, $P_{eq}(x)$ is partially localized on the sites $x_0$ and $L-x_0$ whereas it is almost uniformly distributed over the other sites. It is symmetric with respect to the center of the lattice. At equilibrium,  the average position of the exciton is equal to $\bar{x}=L/2$ and the standard deviation reduces to $\Delta \bar{x}^2=(L^2+2)/12+x_0(x_0/L-1)$. 

In fact, such a behavior can be explained as follows. 
After straightforward calculations, it turns out the zero order RDM in the exciton eigenbasis is expressed as 
\begin{equation}
\sigma(k,k',t) \approx  F(kk',t) e^{-i(\hat{\omega}_{k}-\hat{\omega}_{k'})t}\langle x_0| k'\rangle \langle k | x_0\rangle. 
\label{eq:PX6} 
\end{equation} 
Because the decoherence factor $F(kk',t)$ decays with time provided that $k\neq k'$, the coherences of the RDM gradually disappear. By contrast, the populations are time independent and they remain unchanged. Consequently, the equilibrium corresponds to a statistical mixture of stationary waves $|k\rangle$. The corresponding probabilities, that is,  $|\langle x_0| k\rangle|^2$, define the probabilities to observe the exciton in states $|k\rangle$ given that it occupies the state $|x_0\rangle$. From the expression of the stationary waves (Eq.(\ref{eq:ket})), one easily understands the characteristic of the equilibrium distribution. Note that within the previous scenario, we do not recover that $P_{eq}(x)$ depends on both the temperature and the coupling strength (see Fig. 4). In fact, the influence of these parameters results from the entangled nature of the exciton-phonon eigenstates, entanglement that is neglected within the zero order approximation of the RDM.

\subsection{Linear entropy}

To conclude, let us discuss the behavior of the linear entropy. Because the entropy measures the missing information about the state of the exciton, its time evolution can be interpreted as follows.  At time $t=0$, the exciton is in a pure state $|x_0\rangle$ so that the entropy vanishes. Then, when time elapses, quantum decoherence takes place. The coherences of the RDM gradually disappear so that the exciton is finally described by a statistical mixture of stationary waves. Consequently, this relaxation mechanism is accompanied by an increase of the linear entropy (see Fig. 5). The zero order entropy converges toward a maximum value $S_{\infty}=1-\sum_k|\langle x_0| k\rangle|^4$ defined as
\begin{equation}
S_{\infty}= \left\{ 
\begin{array}{ll}
1-1.5/L & \mbox{if $x_0\neq L/2$} \\
1-2/L & \mbox{if $x_0=L/2$}.
\end{array}
\right.
\end{equation}
However, does the missing information correspond to an information transfer "from the exciton toward the phonon bath" in the form of exciton-phonon correlations? I don't think so. In fact, at zero order, the missing information on the exciton state is due to our ignorance of the phonon states. Indeed, because the phonons are in thermal equilibrium, they are described by a statistical mixture of number states. Therefore, the increase of the entropy originates in the quantum decoherence that results from the average over the phonon degrees of freedom. The missing information on the initial phonon state is thus converted into missing information on the exciton state. Such a conversion is unidirectional so that the entropy behaves as a monotonic function. By contrast, at second order, Fig. 5 has revealed that the linear entropy exhibits high-frequency small-amplitude oscillations. These oscillations account for the fact that the process of information transfer between the exciton and the phonons is partially bidirectional. Basically, a certain amount of information is transferred from the exciton to the phonon bath. Then, owing to the entangled nature of the exciton-phonon eigenstates, a part of this information is restored to the exciton ... and so on. Nevertheless, owing to quantum decoherence, this information exchange gradually disappears and the linear entropy overall increases as time increases. 

\section{Conclusion}

In the present paper, the exciton-phonon problem was revisited for investigating 
exciton-mediated energy redistribution in a finite-size lattice. To proceed, the dynamics was described using the operatorial formulation of PT. Within this method, exciton-phonon entanglement is taken into account through a dual dressing effect, the exciton being dressed by a virtual phonon cloud whereas the phonons are clothed by virtual excitonic transitions. In the nonadiabatic weak-coupling limit, exciton and phonons were treated on an equal footing so that we were able to overcome the problem caused by the fact that the phonons no longer behave as a reservoir.

In that context, special attention has been paid to describing the exciton density that provides information about the way the excitonic energy flows along the lattice. In a marked contrast with what happens in an infinite lattice, we have shown that the dynamics of the density is governed by several time scales. In the short-time limit, the density propagates coherently as if the exciton was insensitive to the phonons. Then, over an intermediate time scale, the exciton-phonon interaction favors quantum decoherence so that the coherent nature of the density gradually disappears. Finally, in the long-time limit, the density converges toward an almost uniform equilibrium distribution. 

From a physical point of view, the exciton density measures the probability for the exciton to tunnel between lattice sites, the phonons evolving freely. Therefore, its behavior results from the quantum interferences between the paths that the exciton can follow to tunnel. The key point is that when the exciton follows two different paths, the phonons develop two different quantum evolutions. The density thus depends on phase factors that reduce to time decaying functions after performing an average over the phonon degrees of freedom. Based on the characterization of the linear entropy, we have shown that the corresponding decay rate is enhanced by the exciton-phonon coupling strength and by the temperature. More surprisingly, our study revealed that an increase of the lattice size softens the influence of the phonons resulting in a slowdown in the decoherence process. Quantum decoherence is thus an intrinsic property of a confined lattice. 

To conclude, we would like to draw the reader's attention to the fact that the present formalism must be restricted to the study of finite-size lattices. Indeed, we have shown that quantum decoherence tends to disappear as the lattice size increases. From these observations, we could extrapolate to conclude that the exciton moves coherently in an infinite lattice. In doing so, we thus recover the results established in Ref.\cite{kn:pouthier3}. Unfortunately, this is a misleading appearance because the origin of the decoherence is twofold. 
First, it results from the modification of the phonon energies induced by the exciton. Consequently, the phonon quantum states evolve differently depending on the state occupied by the exciton. Dynamical phases arise in the exciton RDM so that pure dephasing  occurs when the average over the phonon degrees of freedom is performed (see Eq.(\ref{eq:FACTOR})).
Then, as illustrated by the presence of the unitary transformation $U$ in Eq.(\ref{eq:RDM2}), the decoherence also originates in the entangled nature of the exciton-phonon eigenstates. 
In a confined lattice, the entangled nature of the eigenstates is negligible so that the first origin of the decoherence dominates, as detailed in the present paper. By contrast, in an infinite lattice, the opposite situation occurs because the phonon energy corrections vanish (see Eq.(\ref{eq:dOMpk})). Fortuitously, in the weak-coupling limit, it turns out that the exciton-phonon entanglement does not favor an incoherent (diffusion-like) motion of the exciton so that a coherent (wave-like) regime remains. But this is not always the case. For instance, when one considers the dynamics of the RDM elements that measure the coherence between the vacuum and the one-exciton states, we have shown that quantum decoherence takes place in finite lattices \cite{kn:pouthier6}, owing to exciton-induced phonon energy shift, and in infinite lattices \cite{kn:pouthier8}, owing to the entanglement of the exciton-phonon eigenstates. In that context, forthcoming works will be devoted to establish a quite general approach to treat the exciton-phonon system whatever the lattice size. In doing so, we aim at determining the nature of the decoherence process when the lattice size evolves continuously from a small value to infinity.

\appendix

\section{Generator of the unitary transformation}

As detailed previously \cite{kn:pouthier6}, the generator of the unitary transformation up to second order in $\Delta H$
is given by the following equations 
\begin{eqnarray}
&&\left[ H_A+H_B,S_1 \right] = \Delta H, \nonumber \\
&&\left[ H_A+H_B,S_2 \right] = \frac{1}{2} [ S_1,\Delta H ]^{nd},  
\label{eq:S1S2}
\end{eqnarray}
where $nd$ means the non-diagonal part of an operator in the unperturbed basis. 
From the expression of the exciton-phonon Hamiltonian Eq.(\ref{eq:H}), one easily obtains 
\begin{eqnarray}
S_1&=&\sum_{p} Z_pa_p^{\dag}-Z_p^{\dag} a_p, \nonumber \\
S_2&=&\sum_{pp'} E_{pp'}a_{p'}^{\dag}a_{p}^{\dag}-E_{pp'}^{\dag} a_{p}a_{p'} \nonumber \\
       &+&\sum_{pp'} D_{pp'}a_{p}^{\dag}a_{p'}-D_{pp'}^{\dag} a_{p'}^{\dag}a_{p} +\sum_{p} C_{p}.
\label{eq:S12}
\end{eqnarray}
In the exciton eigenbasis, the various operators that enter the definition of the generator are defined as 
\begin{eqnarray}
\langle k | Z_{p}| k' \rangle &=&\frac{\langle k | M_{p}| k' \rangle}{\omega_k-\omega_{k'}+\Omega_p} \nonumber \\
\langle k | E_{pp'}| k' \rangle&=&\frac{\langle k | B_{pp'}| k' \rangle}{\omega_k-\omega_{k'}+\Omega_p+\Omega_{p'}} \nonumber \\
\langle k | D_{pp'}| k' \rangle&=&\frac{\langle k | \bar{B}_{pp'}| k' \rangle}{\omega_k-\omega_{k'}+\Omega_p-\Omega_{p'}} \nonumber \\
 \langle k | C_{p}| k' \rangle&=&\frac{\langle k | A^{nd}_{p}| k' \rangle}{\omega_k-\omega_{k'}}, 
\label{eq:ZCDE}
\end{eqnarray}
where
\begin{eqnarray}
B_{pp'}&=& \frac{1}{2}[Z_{p},M_{p'}] \nonumber \\
A_{p}&=&-\frac{1}{2}(Z_{p}^{\dag}M_{p}+M_{p}Z_{p}) \nonumber \\
\bar{B}_{pp'}&=&B^{nd}_{pp}\delta_{pp'}+B_{pp'}(1-\delta_{pp'}). 
\label{eq:AB}
\end{eqnarray}

\section{General expression of the exciton RDM}

By inserting the unitary transformation $U$, the coherence Eq.(\ref{eq:RDM}) is rewritten as 
\begin{eqnarray}
\sigma(x,x',t) &=& \label{eq:B1} \\
&&Tr_B \left[ \rho_B \langle x_0 | U^\dag e^{i\hat{H}t} U |x'\rangle  \langle x | U^{\dag} e^{-i\hat{H}t} U  |x_0\rangle \right]. \nonumber
\end{eqnarray}
Since $\hat{H}$ is a sum of independent contributions (see Eq.(\ref{eq:Hhat})), the coherence is expressed as
\begin{eqnarray}
\sigma(x,x',t) &=&\sum_{k=1}^{N} \sum_{k'=1}^{N} \exp\left[-i(\hat{\omega}_k-\hat{\omega}_{k'} )t \right]  \nonumber \\
&&Tr_B \left[ \rho_B  \langle x_0| U^\dag |k' \rangle e^{i\hat{H}_B^{(k')}t}  \langle k'| U|x' \rangle \right. \nonumber \\ 
&& \left. \langle x| U^\dag | k \rangle e^{-i\hat{H}_B^{(k)}t} \langle k| U |x_0 \rangle  
 \right].
 \label{eq:B2}
\end{eqnarray}
Because the operator $\exp(-i\hat{H}_B^{(k)}t)$ defines a unitary evolution and because $[H_B,\hat{H}_B^{(k)}]=0$, the partial trace in Eq.(\ref{eq:B2}) is rewritten as 
\begin{eqnarray}
Tr_B &&\left[ \rho_B  e^{-i(\hat{H}_B^{(k)}-\hat{H}_B^{(k')})t} 
\langle x_0| e^{-i\hat{H}_B^{(k')}t} U^\dag e^{+i\hat{H}_B^{(k')}t} |k' \rangle \right. \nonumber \\
&& \left. \langle k'| U|x' \rangle \langle x| U^\dag | k \rangle  \langle k|e^{-i\hat{H}_B^{(k)}t} U e^{+i\hat{H}_B^{(k)}t} |x_0 \rangle  \right]. \nonumber 
\end{eqnarray}
At this step, let first define $U_k(t)=e^{+i\hat{H}_B^{(k)} t} U e^{-i\hat{H}_B^{(k)} t}$ as the Heisenberg representation of the unitary transformation $U$ with respect to the Hamiltonian $\hat{H}_B^{(k)}$. Then, since $\rho_B=\exp(-\beta H_B)/Z_B$, one obtains
\begin{equation}
\rho_B  e^{-i(\hat{H}_B^{(k)}-\hat{H}_B^{(k')})t}=\frac{Z_B^{(kk')}(t)}{Z_B} \rho_B^{(kk')}(t),
\nonumber 
\end{equation}
with 
\begin{eqnarray}
\rho_B^{(kk')}(t)&=&e^{-\beta H_B-it(\hat{H}_B^{(k)}-\hat{H}_B^{(k')})}/Z_B^{(kk')}(t) \nonumber \\
Z_B^{(kk')}(t)&=&Tr_B \left[ e^{-\beta H_B-it(\hat{H}_B^{(k)}-\hat{H}_B^{(k')})} \right].
\label{eq:B3}
\end{eqnarray}
Finally, inserting Eq.(\ref{eq:B3}) into the previous expression of the partial trace and combining the results with Eq.(\ref{eq:B2}) yield the exciton RDM Eq.(\ref{eq:RDM2}).

\end{document}